\documentclass{ws-rv9x6}
\usepackage{ws-rv-van}             
\makeindex
\begin{document}

\chapter{On the Kolmogorov-Chaitin Complexity for short sequences\label{ch1}}
\author[Jean-Paul Delahaye and Hector Zenil]{Jean-Paul Delahaye\footnote{delahaye@lifl.fr} and Hector Zenil\footnote{hector.zenil-chavez@malix.univ-paris1.fr}}

\address{Laboratoire d'Informatique Fondamentale de Lille\\ Centre National de la Recherche Scientifique (CNRS)\\ Universit\'e des Sciences et Technologies de Lille}

Among the several new ideas and contributions made by Gregory Chaitin to mathematics is his strong belief that mathematicians should transcend the millenary theorem-proof paradigm in favor of a quasi-empirical method based on current and unprecedented access to computational resources\cite{chaitin3}. In accordance with that dictum, we present in this paper an experimental approach for defining and measuring the Kolmogorov-Chaitin complexity, a problem which is known to be quite challenging for short sequences --- shorter for example than typical compiler lengths.

The Kolmogorov-Chaitin complexity (or algorithmic complexity) of a string $s$ is defined as the length of its shortest description $p$ on a universal Turing machine $U$, formally $K(s)=\min\{l(p):U(p)=s\}$. The major drawback of $K$, as measure, is its uncomputability. So in practical applications it must always be approximated by compression algorithms. A string is uncompressible if its shorter description is the original string itself. If a string is uncompressible it is said that the string is random since no patterns were found. Among the $2^n$ different strings of length $n$, it is easy to deduce by a combinatoric argument that one of them will be completely random simply because there will be no enough shorter strings so most of them will have a maximal K-C complexity. Therefore many of them will remain equal or very close to their original size after the compression. Most of them will be therefore random.  An important property of $K$ is that it is nearly independent of the choice of $U$. However, when the strings are short in length, the dependence of $K$ on a particular universal Turing machine $U$ is higher producing arbitrary results. In this paper we will suggest an empirical approach to overcome this difficulty and to obtain a stable definition of the K-C complexity for short sequences.

Using Turing's model of universal computation, Ray Solomonoff\cite{solomonoff,solomonoff2} and Leonid Levin\cite{levin} developed a theory about a universal prior distribution deeply related to the K-C complexity. This work was later known under several titles: universal distribution, algorithmic probability, universal inference, among others\cite{li2,li}. This algorithmic probability is the probability $m(s)$ that a universal Turing machine $U$ produces the string $s$ when provided with an arbitrary input tape. $m(s)$ can be used as a universal sequence predictor that outperforms (in a certain sense) all other predictors\cite{li}. It is easy to see that this distribution is strongly related to the K-C complexity and that once $m(s)$ is determined so is $K(s)$ since the formula $m(s)$ can be written in terms of $K$ as follows $m(s)\approx1/2^{K(s)}$. The distribution of $m(s)$ predicts that non-random looking strings will appear much more often as the result of a uniform random process, which in our experiment is equivalent to running all possible Turing machines and cellular automata of certain small classes according to an acceptable enumeration. By these means, we claim that it might be possible to overcome the problem of defining and measuring the K-C complexity of short sequences. Our proposal consists of measuring the K-C complexity by reconstructing it from scratch basically approximating the algorithmic probability of strings to approximate the K-C complexity.  Particular simple strings are produced with higher probability (i.e. more often produced by the process we will describe below) than particular complex strings, so they have lower complexity.

Our experiment proceeded as follows: We took the Turing machine ($TM$) and cellular automata enumerations defined by Stephen Wolfram\cite{wolfram}. We let run (a) all $2-state$ $2-symbol$ Turing machines, and (b) a statistical sample of the $3-state$ $2-symbol$ ones, both henceforth denoted as $TM(2,2)$ and $TM(3,2)$. 

Then we examine the frequency distribution of these machines' outputs performing experiments modifying several parameters: the number of steps, the length of strings, pseudo-random vs. regular inputs, and the sampling sizes.

For (a) it turns out that there are $4096$ different Turing machines according to the formula $(2sk)^{sk}$ derived from the traditional $5-tuplet$ description of a Turing machine: $d(s_{\{1,2\}}, k_{\{1,2\}})\rightarrow(s_{\{1,2\}}, k_{\{1,2\}}, \{1,-1\})$ where $s_{\{1,2\}}$ are the two possible states, $k_{\{1,2\}}$ are the two possible symbols and the last entry \{1,-1\} denotes the movement of the head either to the right or to the left. From the same formula it follows that for (b) there are $2 985 984$ so we proceeded by statistical methods taking representative samples of size $5 000$, $10 000$, $20 000$ and $100 000$ Turing machines uniformly distributed over $TM(3,2)$. We then let them run $30$, $100$ and $500$ steps each and we proceeded to feed each one with (1) a (pseudo) random (one per TM) input and (2) with a regular input.

We proceeded in the same fashion for all one dimensional binary cellular automata ($CA$), those (1) which their rule depends only on the left and right neighbors and those considering two left and one right neighbor, henceforth denoted by $CA(t,c)$\footnote{A better notation is the $3-tuplet$ $CA(t,c,j)$ with $j$ indicating the number of symbols, but because we are only considering $2-symbol$ cellular automata we can take it for granted and avoid that complication.} where $t$ and $c$ are the neighbor cells in question, to the left and to the right respectively. These CA were fed with a single $1$ surrounded by $0$s. There are $256$ one dimensional nearest-neighbor cellular automata or $CA(1,1)$, also called Elementary Cellular Automata\cite{wolfram}) and $65536$ $CA(2,1)$.

To determine the output of the Turing machines we look at the string consisting of all parts of the tape reached by the head.  We then partition the output in substrings of length k. For instance, if k=3 and the Turing machine head reached positions 1, 2, 3, 4 and 5 and the tape contains the symbols \{0,0,0,1,1\} then we increment the counter of the substrings $000$, $001$, $011$ by one each one. Similar for $CA$ using the "light cone" of all positions reachable from the initial 1 in the time run. Then we perform the above for (1) each different $TM$ and (2) each different $CA$, giving two distributions over strings of a  given length $k$.

We then looked at the frequency distribution of the outputs of both classes $TM$ and $CA$\footnote{Both enumeration schemes are implemented in Mathematica calling the functions CelullarAutomaton and TuringMachine, the latter implemented in Mathematica version $6.0$}, (including ECA) performing experiments modifying several parameters: the number of steps, the length of strings, (pseudo) random vs. regular inputs, and the sampling sizes. 


An important result is that the frequency distribution was very stable under the several variations described above allowing to define a  \emph{natural} distribution $m(s)$ particularly for the top it. We claim that the bottom of the distribution, and therefore all of it, will tend to stabilize by taking bigger samples. By analyzing the following diagram it can be deduced that the output frequency distribution of each of the independent systems of computation (TM and CA) follow an output frequency distribution. We conjecture that these systems of computation and others of equivalent computational power converge toward a single distribution when bigger samples are taken by allowing a greater number of steps and/or bigger classes containing more and increasingly sophisticated computational devices. Such distributions should then match the value of $m(s)$ and therefore $K(s)$ by means of the convergence of what we call their experimental counterparts $m_e(s)$ and $K_e(s)$. If our method succeeds as we claim it could be possible to give a stable definition of the K-C complexity for short sequences independent of any constant.
\\

\begin{figure}
\label{TMgraph1}
\begin{center}
\includegraphics[height=3.055in, width=4.5in]{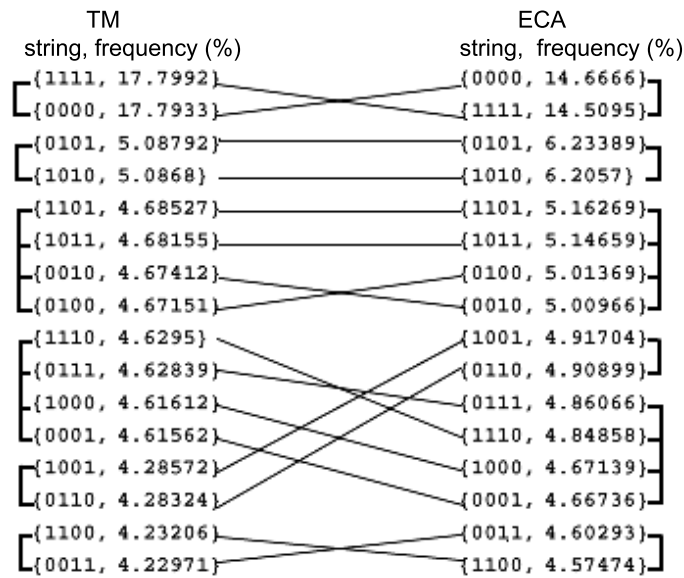}
\end{center}
\caption{The above diagram shows the convergence of the frequency distributions of the outputs of $TM$ and $ECA$ for $k=4$. Matching strings are linked by a line. As one can observe, in spite of certain crossings, $TM$ and $ECA$ are strongly correlated and both successfully group equivalent output strings. By taking the six groups --- marked with brackets --- the distribution frequencies only differ by one.}
\end{figure}

By instance, the strings $0101$ and  $1010$ were grouped in second place, therefore they are the second most complex group after the group composed by the strings of a sequence of zeros or ones but before all the other $2^n$ strings. And that is what one would expect since it has a very low K-C complexity as prefix of a highly compressible string $0101\ldots$. In favor of our claims about the nature of these distributions as following $m(s)$ and then approaching $K(s)$, notice that all strings were correctly grouped with their equivalent category of complexity under the three possible operations/symmetries preserving their K-C complexity, namely reversion $(sy)$, complementation $(co)$ and composition of the two $(syco)$. This also supports our claim that our procedure is working correctly since it groups all strings by their complexity class. The fact that the method groups all the strings by their complexity category allowed us to apply a well-known lemma used in group theory to enumerate actual different cases, which let us present a single representative string for each complexity category. So instead of presenting a distribution with $1024$ strings of length $10$ it allows us to compress it to $272$ strings.

We have also found that the frequency distribution from several real-world data sources also approximates the same distribution, suggesting that they probably come from the same kind of computation, supporting contemporary claims about nature as performing computations\cite{wolfram,lloyd}. The paper available online contains more detailed results for strings of length $k=4,5,6,10$ as well as two metrics for measuring the convergence of $TM(2,2)$ and $ECA(1,1)$ and the real-world data frequency distributions extracted from several sources\footnote{It can be reached at arXiv: http://arxiv.org/abs/0704.1043.\\ A website with the complete results of the whole experiment is available at http://www.mathrix.org/experimentalAIT/}. A paper with mathematical formulations and further precise conjectures is currently in preparation.

\end{document}